\documentclass[useAMS,usenatbib]{mn2e}
\usepackage{psfig}
\usepackage{ltxtable}

\newcommand{\myemail}{yaoys@astro.umass.edu}
\newcommand{\etal}{et al.~}
\def\gsim{\lower 2pt \hbox{$\, \buildrel {\scriptstyle >}\over
{\scriptstyle \sim}\,$}}
\def\lsim{\lower 2pt \hbox{$\, \buildrel {\scriptstyle <}\over
{\scriptstyle \sim}\,$}}

\title[Black Hole X-ray Binary LMC~X--1 \& LMC~X--3]{Black Hole X-ray Binary LMC~X--1 and LMC~X--3: Observations Confront Spectral Models}
\author[Yao, Wang, \& Zhang]{Yangsen Yao$^{1}$\thanks{E-mail:\myemail}, 
Q. Daniel Wang$^1$, and Shuang Nan Zhang$^{2}$\\
$^{1}$Department of Astronomy, University of Massachusetts, Amherst, MA 01003;\\
$^{2}$Department of Physics, University of Alabama, Huntsville, AL 35899}
\begin{document}

\date{Accepted 2005 xxxx xx. Received 2005 January xx; in original form 2005 xxxx xx}

\pagerange{\pageref{firstpage}--\pageref{lastpage}} \pubyear{2002}

\maketitle

\label{firstpage}

\begin{abstract}
We present a comprehensive spectral analysis of black hole X-ray binaries,
LMC~X--1 and LMC~X--3, based on {\sl BeppoSAX} observations. We test both 
the multi-color disk plus power law (MCD+PL) model and a newly-developed 
Monte-Carlo simulation-based model for a Comptonized MCD
(CMCD) with either a spherical or a slab-like corona, 
by comparing the inferred 
parameters with independent direct measurements. While all models give 
an adequate description of the spectra, we find a
significant discrepancy between the MCD+PL
inferred X-ray-absorbing gas column density and the absorption-edge
measurement based on dispersed X-ray spectra. The MCD+PL
fits to the LMC X--1 spectra 
also require a change in the inner disk radius during the 
{\sl BeppoSAX} observation, which may be due to the nonphysical 
effects inherited in the model. In contrast, the CMCD model with the spheric 
corona gives the predictions of both the disk inclination angle and
the absorption that are consistent with the direct measurements, 
and only slightly under-predicts the black hole mass of LMC~X--3.
The model explains the spectral state evolution 
of LMC~X--1 within the {\sl BeppoSAX} observation as a change
in the accretion rate, which leads to an increase in both the inner disk
temperature and the Comptonization opacity. On the other hand,
the CMCD model with the slab-like corona is more problematic in the test
and is thus not recommended. 

\end{abstract}


\section{Introduction}
Accreting black hole X-ray binaries (BHXBs) typically stay in one of the two 
distinct spectral states \citep{tanaka95}. In the so-called high/soft 
state, such a binary has a soft X-ray spectrum and a relatively high 
X-ray luminosity, which is believed to be dominated by the emission 
directly from the accretion disk around the black hole (BH). 
Under the standard 
geometrically thin and optically thick accretion disk approximation 
(see Pringle 1981 and references therein),
the X-ray spectrum is an integration of the assumed blackbody-like emission 
over the disk with a temperature that decreases with increasing 
radius \citep{mit84} 
and is thus called the multi-color disk (MCD). In the opposite 
low/hard state, the spectrum is relatively flat and can often be 
approximated by a power law (PL), which may extend up to several 
hundred keV, whereas the luminosity is typically low.
The flat spectral shape is 
usually attributed to the Comptonization of soft
disk photons by hot electrons in a surrounding corona (e.g., Zdziarski 2000).
Therefore the spectrum of a BHXB system can usually be well fitted by 
a two-component model with a blackbody-like (soft) component plus
a PL-like (hard) component;  a disk reflection component and 
a broad iron line component are sometimes also visible in 
the spectra of some systems (e.g., Cygnus X-1, Di Salvo \etal 2001; 
GX 339--4, Ueda, Ebisawa, \& Done 1994).

Of course, there are various other effects that one needs to consider.
For the soft component, 
because of the high temperature ($\sim$ 1 keV) in the inner region of the 
accretion disk, disk emission is expected to be slightly Comptonized 
by the free electrons in this region and the local emergent spectrum 
could be approximated as a diluted blackbody spectrum rather than a blackbody 
one. Furthermore, the emission from the inner accretion disk is 
subject to the strong gravitational field in the vicinity of the BH, 
and should also be modified by the extreme Doppler motion of the 
accretion disk. 
Detailed calculations suggest that the simple MCD model (diskbb in XSPEC)
could still be used to describe the distorted disk emission, but needs 
to be corrected by several factors in order to infer the consistent 
physical quantities: a factor $f_{color}$ accounting for the 
color temperature correction of the 
Comptonization in the inner disk region (e.g, Shimura \& Takahara 1995;
Merloni, Fabian, \& Ross 2000);
a factor $\eta$ indicating the difference between the apparent and
intrinsic radii of the peak temperature, and the integrated luminosity
difference between the MCD model and a more realistic accretion disk
with torque-free boundary condition 
(e.g., Kubota \etal 1998; Gierli\'{n}ski \etal 1999);
factors $f_{GR}$ and $g_{GR}$ accounting for gravitational redshift and 
Doppler shift and the integrated disk flux change due to the general 
relativistic effects (e.g., Cunningham 1975; also see Zhang, Cui, \& Chen
1997 for more discussions).

For the hard component, the phenomenological PL model is often used. However, 
the extrapolation of the PL model to the lower energy in the MCD+PL model
neglects the seed photon curvature
that should be reflected in the Comptonized spectrum 
(e.g., Shrader \& Titarchuk 1998; Done, \.{Z}ycki, \& Smith 2002).
In a typical global fit of the MCD+PL model to a BHXB spectrum, 
this unphysical extrapolation usually leads to 
an artificial increase in the soft X-ray-absorbing
column density estimates. As a result, the  absorption-corrected source 
flux could be over-estimated. This problem is not very acute for 
an observation with poor counting statistics and/or for a
low temperature accretion disk with the MCD emission peak located 
outside the observed energy range (e.g., accreting intermediate-mass BH
candidates; Wang \etal 2004, Paper II). But for a stellar mass BH with an
accretion disk temperature of $\sim 1$ keV, the problem is 
significant. In fact, it has been shown that the MCD+PL model fails
to give an acceptable fit to X-ray spectra of LMC X--3, when  
the X-ray absorption is tightly constrained by an independent measurement
from X-ray absorption edges in dispersed X-ray spectra \citep{pag03}.

Instead of using the PL model, several groups
have developed various Comptonization models by taking care of 
the radiative transfer process either numerically or analytically
(e.g., $compbb$, Nishimura, Mitsuda, \& Itoh 1986; $comptt$, Titarchuk 1994; 
Hua \& Titarchuk 1995; $thcomp$, \.{Z}ycki, Done, \& Smith 1999; 
$eqpair$, Coppi 1999 and references therein; 
$compps$, Poutanen \& Svensson 1996; etc.). 
The seed photon spectrum is assumed to be a single-temperature
blackbody in $compbb$ or its Wien approximation in $comptt$,
which clearly deviates from a multi-color black-body disk.
This deviation could be very significant, especially for a disk with a high
inner temperature, typical for a stellar mass BH; the black-body 
would peak well within an X-ray band. 
Whereas the $thcomp$ is a thermal Comptonization model, the $eqpair$ model
is a hybrid model of thermal and non-thermal Comptonization
(see also Gierli\'{n}ski \etal 1999); both models
can adopt a proper disk spectrum, but the latter is more advanced
which takes care of nearly all the
important physical processes in a disk-corona system including Compton
scattering, pair production and annihilation, bremsstrahlung, and 
synchrotron radiation, etc. \citep{coppi92}. All these four models can 
only produce a direction-averaged spectrum. Like $eqpair$ model, 
the $compps$ model also contains most physical processes
in the disk-corona system and allows the use of a MCD spectrum, but it
can treat geometry more accurately and generate an angular dependent 
spectrum in a spherical or a slab-like corona system.

It is worth noting that in the disk-corona scenario, 
the Comptonization generates the PL-like component at the expense of 
the MCD flux, therefore the MCD and the PL-like components are physically 
related. Some Comptonization models mentioned above 
(e.g., $eqpair$ and $compps$) have taken care of this relationship 
whereas some have not (e.g., $thcomp$). 
One then should be cautious of interpreting the physical meaning of the 
MCD parameters when applying the Comptonization models. 
The MCD model here describes only those un-Comptonized (un-scattered 
or escaped) disk photons rather than the actually original disk emission. 
In particular, this relationship does not exist in MCD+PL model, therefore
the original disk flux could be under-estimated from the fitted MCD 
parameters, so are other related inferred parameters such as inner disk radius.
The under-estimate, for example, may be 
responsible for the {\sl apparent}
change in the inner disk radius with the transition from one state to another,
as has been claimed for several sources such as
XTE J1550-564, GRO J1655-40 \citep{sob99a, sob99b}, and
XTE J2012+381 \citep{cam02}.

All the models mentioned above 
have been widely used to model the spectra of Galactic BHXBs, 
neutron star X-ray binaries, as well as the extragalactic sources
including ultraluminous X-ray sources (ULXs) and AGNs.
However, except for the works by \citet{gie99,gie01}, 
there has been little rigorous test of the models
against direct measurements (e.g., neutral absorption
column density, BH mass, system inclination, etc.), which are 
available for several well-known 
systems such as GRO~J1655--40, LMC~X--1, and LMC~X--3.

In Yao \etal (2005; Paper I), we have presented a Monte-Carlo method in 
simulating 
Comptonized multi-color disk (CMCD) spectra. The simulations used the MCD as 
the source of seed photons and self-consistently accounted for the radiation 
transfer in the Comptonization in a spherical or slab-like thermal plasma. 
We have applied this CMCD model, implemented as a
table model in XSPEC, to a stellar mass BH 
candidate XTE~J2012+381 in our Galaxy. This application
shows that the inner disk radius is {\sl not} required to change
when the source transits from the soft state to the hard state, 
in contrast to the conclusion reached from the fits with the 
MCD+PL model \citep{cam02}. 
For a spherical corona, the toy model contains the following parameters:
the inner disk temperature ($T_{in}$), the system inclination 
angle ($\theta$), 
the effective thermal electron temperature ($T_c$), optical depth ($\tau$), 
and radius ($R_c$) of the corona as well as the normalization 
defined as
\begin{equation} \label{equ:cmcd}
K_{CMCD} = \left(\frac{R_{in}/\mathrm{km}}{D/10\mathrm{kpc}}\right)^2,
\end{equation}
where $D$ is the source distance and $R_{in}$  is the apparent inner
disk radius. For a slab-like corona, another extreme of the geometry,
we assume that the corona covers 
the whole accretion disk. We find that the final emerged spectra are 
insensitive to the different vertical scales, 
so $R_c$ does not appear as a parameter in this geometry. 
Note that $ K_{CMCD}$ differs from the normalization 
for the MCD model, % (e.g., {\sl diskbb} in XSPEC),
\begin{equation} \label{equ:mcd}
K_{MCD} = \left(\frac{R_{in}/\mathrm{km}}{D/10\mathrm{kpc}}\right)^2\cos(\theta). 
\end{equation}
More detailed discussion on the CMCD model 
can be found in Paper II, in which we have applied the same model for a 
spherical geometry to six ULXs
observed with {\sl XMM-Newton}. The fitted $T_{in}$
($\sim$ 0.05--0.3 keV) of these sources 
are distinctly different from the values ($\sim 1$ keV)
obtained for known stellar-mass BHs, as presented in this paper and in 
Paper I. Indeed, the inferred BH masses ($M_{BH}$)
of the ULXs are $\sim 10^3 M_\odot$, consistent with the intermediate-mass
BH interpretation of these sources. We have also shown that the MCD+PL model
gives an equivalent spectral description of 
the ULXs, although the CMCD model provides unique constrains on the 
corona properties and on the disk inclination angles, as well as on
the BH masses.
Because of the lower disk temperatures, compared to those of the
stellar mass BH systems, the nonphysical
effects of the MCD+PL model are typically not significant in the observable 
photon energy range of the intermediate-mass
BH candidates. 

In the present work, we conduct a critical test of the CMCD and
MCD+PL models by comparing parameters ($M_{BH}$, $\theta$, and 
the equivalent neutral hydrogen absorption $N_H$)
inferred from the X-ray spectra of LMC X--1 and
X--3 with the more direct measurements based on optical and dispersed X-ray
spectra; we also compare the results from the different corona geometrical
configurations of the CMCD model. We first briefly describe these 
measurements and the X-ray observations in \S 2, and then 
present the spectral fitting results in \S 3. We describe the specific
comparisons in \S 4 and present the discussion and our conclusions in \S 5.

\section{Description of the Sources and Observations}

We select LMC~X--1 and X--3 for this study chiefly because of their 
location in our nearest neighboring galaxy,
the Large Magellanic Cloud (LMC; $D = 50$ kpc is adopted throughout the
work). Both the well-determined distance and the relatively low foreground
soft X-ray absorption are essential to our test. These two sources are also
among the three well-known persistent BHXBs and are usually found 
in the high/soft state. The low/hard state was  occasionally reported
for LMC~X--3 \citep{boy00, hom00}, but never for LMC~X--1. 
The remaining known persistent BHXB, Cygnus~X--1, is 
in our Galaxy and stays mostly in the low/hard state
(e.g., Pottschmidt \etal 2003).  The X-ray spectra of this source
also show a strong disk reflection component \citep{gil99, fro01} --- 
a complication that is not included in the CMCD models. 
Table~\ref{tab:keymeasure} summarizes the key parameters of LMC~X--1 and 
X--3,
which are used for the comparison with our spectrally inferred values (\S 4).

\begin{table*}
  \centering
  \begin{minipage}{140mm}
    \caption{Comparison of parameter measurements. \label{tab:keymeasure}}
    \begin{tabular}{@{}l|ccccl@{}}
      \hline
      & {$\theta$}   & {M$_{BH}$}   & {N$_H$}                & {T$_{in}$ } & \\
      & {($^\circ$)}  & {(M$_\odot$)}& {(10$^{20}$cm$^{-2}$)} & {(keV)}      & References\\
      \hline
      & \multicolumn{4}{c}{\underline{LMC~X--1}}\\
      Indep. est.    & 24$\le\theta\le$64        & 4$\le$M$\le$12.5 &--&--& 1, 2\\
      CMCD: sphere   & $\lsim$43   & 4.0(3.8--4.5)    & 54(52--56) & 0.93(0.91--0.94)\\
      CMCD: slab     & $\gsim$64   & 15.2(10.4--19.9) & 53(51--54) & 0.80(0.79--0.81)\\
      MCD+PL         &  N/A      & 3.0(2.9--3.1)    & 84(79--89) & 0.93(0.92--0.95)\\
      \hline
      & \multicolumn{4}{c}{\underline{LMC~X--3}}\\
      Indep. est.    & $\theta\le$70 & M$>5.8\pm0.8$ & 3.8(3.1--4.6) &-- & 3,4,5,6\\
      CMCD: sphere   & $\lsim$69 & 4.19(4.17--4.21)    & 4.5(4.2--4.7) & 0.98(0.97--0.99)\\
      CMCD: slab     & $\lsim$61 & 3.73(3.71--3.76)    & 4.4(4.2--4.6) & 0.96(0.95--0.97)\\
      MCD+PL         &  N/A    & 4.2(4.1--4.3)       & 7.0(6.0--8.0) & 1.02(1.01--1.03)\\
      \hline
    \end{tabular} \\
    {The BH mass M$_{BH}$ is estimated by assuming zero spin of the BH.
      See text for details.
    References: 
    $^1$ \citet{hut83}; 
    $^2$ \citet{gie01};
    $^3$ \citet{cow83};
    $^4$ \citet{pac83};
    $^5$ \citet{sor01};
    $^6$ \citet{pag03}.}
  %\end{list}
  \end{minipage}
\end{table*}

Both LMC X--1 and X--3 were observed with {\sl Chandra} 
(e.g., Cui \etal 2002) and {\sl XMM-Newton} (e.g., Page \etal 2003). 
The dispersed X-ray spectra of LMC~X--3 have been used to measure the
X-ray absorption edges (mainly for Oxygen), which tightly constrains the
the absorbing matter column density $N_H$ along the line of sight 
\citep{pag03}. The absorption towards LMC X--1 is, however, substantially
higher. As a result, the photon flux at the Oxygen edge is too low to 
allow for a useful constraint on $N_H$ based on the existing data. 

We here utilize the data from the {\sl BeppoSAX} observations, which 
were carried out on 1997 October 5 for LMC X--1 and 
October 11 for X--3 \citep{tre00}. The data do not have pile-up problems,
which could be present for X-ray CCD imaging observations of bright sources.
Four types of narrow-field instruments (NFIs) were on board: 
Low Energy Concentrator System (LECS), Medium Energy Concentrator 
Systems (MECS), High Pressure Gas Scintillation Proportional 
counter (HPGSPC), and Phoswich Detector System (PDS) \citep{boella97}.
The exposure for LMC~X--1 and X--3 were about 15 ks each for the LECS and 
about 40 ks each for the MECS. These two instruments were sensitive to X-rays 
in the energy ranges of 0.1--10 and 1.3--10 keV respectively.
Data from the HPGSPC and the PDS, which were sensitive to photons in 4--120 
keV and 15--300 keV ranges respectively, were not included because of 
poor counting statistics, and also because of possible source contamination 
from PSR 0540--69 (which is 25$^\prime$ away from 
LMC~X--1) \citep{sew84,haa01}. We extracted the spectra from a radius 
of 8$'$ and 8.4$'$ around each source
from the LECS and the MECS observations and used the energy ranges of
0.2--4 keV and 1.8--10 keV for these two instruments in this study. 
The background contributions to the spectra are small and are estimated from 
a blank field. 
The spectra from the LECS, the MECS2, and the MECS3 were jointly fitted 
for each source, using the software package XSPEC{\em 11.2.0bs}.

\section{Results}

\begin{figure}
\centering
\psfig{figure=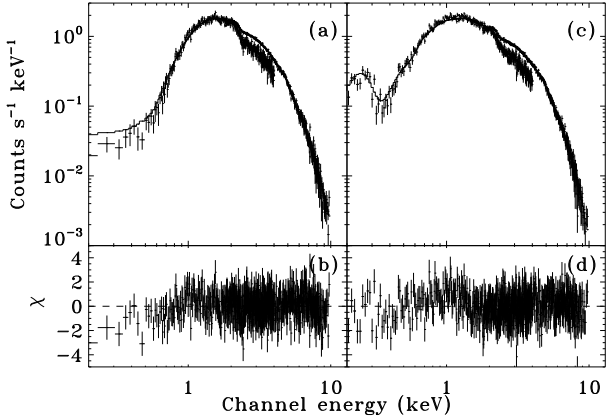,width=0.45\textwidth}
\caption{Model fits to the {\sl BeppoSAX} spectra of LMC~X--1 
({\sl panel a}) and LMC~X--3 ({\sl panel c}), and the corresponding 
residuals in term of sigmas ({\sl panels b} and {\sl d}).
The {\sl solid line} in {\sl panels a} and {\sl c} show the fit of CMCD model.
The fit goodnesses from the different geometrical models (a sphere vs. a slab) 
are nearly the same.
\label{fig:cmcd}}
\end{figure}

\begin{figure}
\centering
\psfig{figure=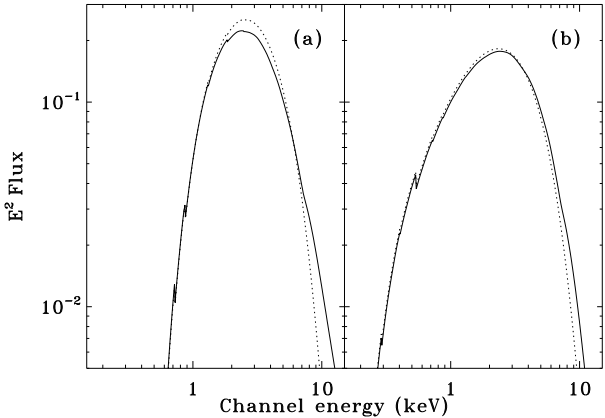,width=0.45\textwidth} 
\caption{The effects of the Comptonization in LMC~X--1 ({\sl a}) and in
LMC~X--3 ({\sl b}). {\sl Solid line}: spherical CMCD model with the best
fit parameters; {\sl dotted line}: CMCD model with the same parameters but 
$\tau$=0, which is equivalent to the MCD model.
\label{fig:compton}}
\end{figure}

We summarize the spectral fitting results and the 
inferred source fluxes in Table~\ref{tab:fit-parameters}. 
The quoted uncertainty ranges of the parameters are all at 90\% 
confidence level. Fig.~\ref{fig:cmcd} shows the spectral fits with the 
CMCD models. The systematic deviation of the data from the model at 
low energies
($\la 1$ keV) might be due to poor calibration
of the instrument spectral response \citep{martin96}.
Fig.~\ref{fig:compton} illustrates the effects of the Comptonization in the
spherical corona systems.

\begin{table}
%  \centering
  \caption{Spectral fit results \label{tab:fit-parameters}}
  \begin{tabular}{@{}lll@{}}
    \hline 
    {parameters} & {LMC X--1} & {LMC X--3}\\
    \hline
    \multicolumn{3}{c}{\underline{CMCD: sphere}} \\ %\sidehead{CMCD}
    $N_H$ (10$^{21}$ cm$^2$) &  5.4(5.2--5.6)    & 0.45(0.42--0.47) \\ 
    $T_{in}$ (keV)           &  0.93(0.91--0.94) & 0.98(0.97--0.99) \\
    $T_c$    (keV)           &  19(15--23)       & 20(19--23)       \\
    $R_c$    (R$_g$)         &  11(9--19)        & 10.0(9.6--16.6)  \\ 
    $\tau$                   &  1.0(0.8--1.3)    & 0.10(0.09--0.13) \\ 
    $\theta$ (deg)           &  28(**--43)       & 59(**--69)       \\ 
    $K_{\rm CMCD}$           &  57(51--73)       & 44.0(43.5--44.3) \\
    $\chi^2/dof$             &  641/627          & 760/649          \\
    $f_{0.2-10}$             &  8.6              & 5.7    \\
    \hline
    \multicolumn{3}{c}{\underline{CMCD: slab}} \\ %\sidehead{CMCD}
    $N_H$ (10$^{21}$ cm$^2$) &  5.3(5.1--5.4)    & 0.44(0.42--0.46) \\ 
    $T_{in}$ (keV)           &  0.80(0.79--0.81) & 0.96(0.95--0.97) \\
    $T_c$    (keV)           &  10.0(9.0--10.6)  & 22(19--24)        \\
    $\tau$                   &  0.45(0.41--0.49) & 0.09(0.08--0.12) \\
    $\theta$ (deg)           &  75(64--**)       & 50(**--61)       \\
    $K_{\rm CMCD}$           &  489(230--840)    & 42.7(42.1-- 43.3)\\
    $\chi^2/dof$             &  638/628          & 759/650          \\
    $f_{0.2-10}$             &  8.5              & 5.7              \\
    \hline
    \multicolumn{3}{c}{\underline{MCD+PL}} \\ %\sidehead{CMCD}
    %% \cutinhead{diskbb + powerlaw}
    $N_H$ (10$^{21}$ cm$^2$) & 8.4(7.9--8.9) & 0.7(0.6--0.8)        \\
    $T_{in}$ (keV)           & 0.93(0.92--0.95) & 1.02(1.01--1.03)  \\
    $K_{\rm MCD}$            & 28(26--30)       & 23(22--24)     \\
    $\Gamma$                 & 3.5(3.4--3.6)    & 2.6(2.5--2.8)     \\
    $K_{PL}$  (10$^{-1}$)    & 2.3(1.9--2.7)    & 0.19(0.14--0.23)\\
    $\chi^2/dof$             & 624/629          & 747/651            \\
    $f_{0.2-10}$             & 31               & 6.4               \\
    \hline
    \end{tabular}\\
{The uncertainty ranges are given in parenthesis at the 90\% 
confidence level; asterisks indicate that the limit is not constrained.
$R_g$ = GM/c$^2$. The $f_{0.2-10}$ is the absorption-corrected 
flux in the energy range 0.2--10 keV and in unit of 
$10^{-10}{\rm~erg~cm^{-2}~s^{-1}}$.}
\end{table}

Both CMCD and MCD+PL models give acceptable fits. 
The model parameters are all well constrained except for the 
CMCD parameter $\theta$, for which only the upper 
or lower limit is constrained.
The MCD+PL model parameters we obtained here 
are consistent with those reported by \citet{haa01}.
The fitted $N_H$ values from MCD+PL are systematically 
higher than those from CMCDs. The same is true for the inferred 
absorption-corrected fluxes, especially for LMC~X--1 which is a factor of 
$\sim$ 4 higher from MCD+PL than from CMCD (Table~\ref{tab:fit-parameters}).

For LMC~X--3, $T_{in}$ from MCD+PL is slightly higher than those from CMCDs,
whereas the value of $K_{MCD}/{\rm cos}(\theta)$ is consistent with those of
$K_{CMCD}$. The best fit parameters in the two different geometric CMCDs  
are nearly identical. The small value of $\tau_c$ indicates that only
a small portion of disk photons have been up-scattered to high energies.

For LMC~X--1, except for a consistent $N_H$ value, the fitted parameters 
are significantly different 
between the two different geometric CMCDs (Table~\ref{tab:fit-parameters}).
We will see in \S4, the results from the slab-like configuration 
are inconsistent with the independent measurements.
$T_{in}$ values from the spherical CMCD and the MCD+PL 
are consistent with each other, but $K_{MCD}/{\rm cos}(\theta)$ value, 
assuming the best-fit CMCD $\theta$, is $\sim$ 1.5 smaller than $K_{CMCD}$.

\begin{figure}
\centering
\psfig{figure=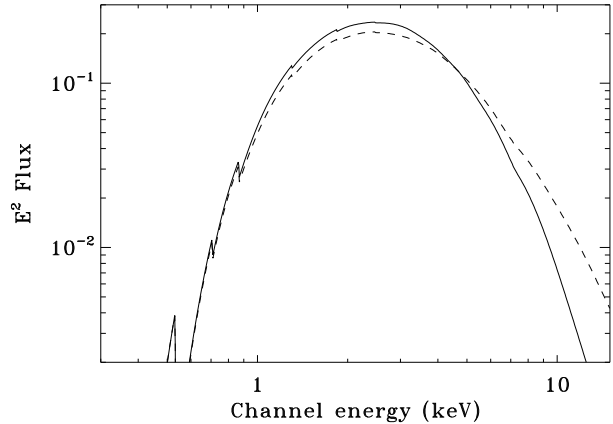,width=0.45\textwidth} 
\caption{The spectral variation of LMC~X--1 from the first 30 ks (part 1, 
{\sl solid line}) to the remaining time (part 2, {\sl dashed line}). 
The models are plotted with the best fitting parameters
(Table~\ref{tab:evolution}). 
\label{fig:evolution}}
\end{figure}

Because a significant change in the spectral shape occurred
during the observation of LMC X--1, we have further
split the exposure into two parts, the first 30 ks and 
the remaining time, in the same way as in \citet{haa01}.
To tighten the constraints on spectral parameters,
we jointly fit $N_H$ and $\theta$, which should
be the same in the two parts of the observation (Table~\ref{tab:evolution}).
From the early part to the later part, according to the 
spherical CMCD model,
$T_{in}$ increased by $\sim$ 7\% 
and the corona became a factor of $\sim$ 2 larger and
opaque with $\tau$ increased by a factor of $\sim$ 4. The source flux
in the 0.2-10 keV band decreased slightly, although the normalization 
remained essentially the same. Because 
$T_{in} \propto (\dot{M}/M_{BH})^{1/4}$, where $M_{BH}$ and $\dot{M}$ are
the BH mass and the accretion mass rate, a rising $T_{in}$ during
the observation of LMC X--1 was then caused by an increasing 
accretion rate. Apparently, this change led to the thickening of the 
corona. The slight decline of the 
flux is likely due to a combination of the energy loss to the
corona and to the scattering of photons to energies greater than 
$\ga 10$ keV. Fig.~\ref{fig:evolution} demonstrates the differences
of the Comptonization effects in the two parts of the observation. 
However, it is hard to physically understand the fitted parameters of the 
slab-like CMCD: From early part to the later part, 
$T_{in}$ and $T_c$ need to decrease by $\sim16$\% and a factor of
4, respectively, but in the mean time both $\tau$ and $K_{CMCD}$ 
(hence $R_{in}$) have to increase dramatically. 
We believe that the slab-like CMCD model is not suitable 
for describing such a spectral state of a BHXB system when
the hard component contributes significantly to the total flux.

\begin{table}
\caption{Spectral variation of LMC X--1 \label{tab:evolution}}
\begin{tabular}{@{}lll@{}}
\hline
{parameters} & {part 1} & {part 2} \\
\hline
\multicolumn{3}{c}{\underline{CMCD: sphere}}\\
$N_H$ (10$^{21}$ cm$^2$) &  5.3(5.2--5.5)      & = part 1        \\
$T_{in}$ (keV)           &  0.91(0.90--0.92)   & 0.96(0.93--0.99)\\
$T_c$    (keV)           &  13(9--18)          & 10(8--12)       \\
$R_c$    (R$_g$)         &  10(9--16)          & 25(18--33)      \\ 
$\tau$                   &  0.55(0.5--0.7)     & 2.0(1.8--2.4)   \\
$\theta$ (deg)           &  23(**--45)         & = part 1        \\ 
$K_{CMCD}$               &  61(55--72)         & 57(46--77)      \\
$\chi^2/dof$             &  \multicolumn{2}{c}{1211/1192}       \\
$f_{0.2-10}$             &  8.8                & 8.3             \\ 
%$f_{0.2-10}$             &  8.8(7.9--10.4)     & 8.3(6.7--11.2)  \\ 
\hline
\multicolumn{3}{c}{\underline{CMCD: slab}}\\
$N_H$ (10$^{21}$ cm$^2$) &  5.3(5.2--5.4)      & = part 1        \\
$T_{in}$ (keV)           &  0.81(0.80--0.82)   & 0.74(0.73--0.76)\\
$T_c$    (keV)           & 20(19--23)         & 5(**--7)     \\
$\tau$                   &  0.11(0.10--0.14)   & 1.9(1.7--2.1)   \\
$\theta$ (deg)           &  75(64--**)         & = part 1        \\ 
$K_{CMCD}$               &  348(224--495)     & 1223(730--1855)   \\
$\chi^2/dof$             & \multicolumn{2}{c}{1209/1194}          \\
$f_{0.2-10}$             &  8.8                & 8.1             \\ 
\hline                    
\multicolumn{3}{c}{\underline{MCD+PL}}  \\ %{MCD+PL} \\
$N_H$ (10$^{21}$ cm$^2$) & 8.5(8.0--9.0)& = part 1             \\
$T_{in}$ (keV)           & 0.90(0.89--0.91) & 1.00(0.98--1.03) \\
$K_{MCD}$                & 52(49--55)       & 23(21--26)       \\
$\Gamma$                 & 3.6(3.5--3.8)    & 3.4(3.3--3.6)    \\
$K_{PL}$  (10$^{-1}$)      & 3.0(2.4--3.7)    & 3.4(2.8--3.9)  \\
$\chi^2/dof$             & \multicolumn{2}{c}{1180/1195}      \\
$f_{0.2-10}$             &  48              & 42               \\ 
%$f_{0.2-10}$             &  48(38--57)      & 42(35--49)       \\ 
\hline
\end{tabular} \\
{Please refer to Table~\ref{tab:fit-parameters}.}
\end{table}

Now let us check the results from the MCD+PL model, which 
are fairly consistent with those of \citet{haa01}. The MCD normalizations are 
significantly different between the early and later
parts of the observation (Table~\ref{tab:evolution}), confirming
the analysis by \citet{haa01}, which is based on the same data and
same model. They concluded that as $T_{in}$ increased (by $\simeq$ 9\%), 
$R_{in}$ decreased ($\simeq$ 38\%) from the 
early to later parts of the observation. 
But, as was discussed in \S 1, this apparent change in $R_{in}$ 
is most likely due to 
the lack of the accounting for the radiative transfer between
the two components of the MCD +PL model and is therefore not physical.

\begin{figure}
\centering
\psfig{figure=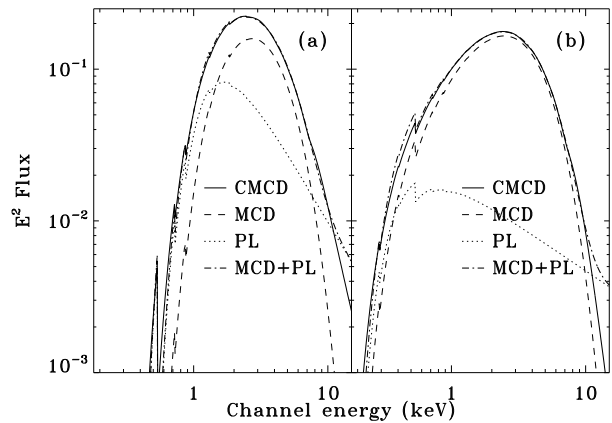,width=0.45\textwidth}
\caption{The comparison of the spherical CMCD model and MCD+PL model with the best fit 
parameters (Table~\ref{tab:fit-parameters})
for ({\sl a}) LMC~X--1 and ({\sl b}) LMC~X--3.
\label{fig:compare}}
\end{figure}

Furthermore, Fig.~\ref{fig:compare} shows clearly that
for both sources, the PL component surpasses the MCD component in
contributing to the spectra at low
energies ($\lsim$ 1 keV for LMC~X--1 and $\lsim$ 0.4 keV for LMC~X--3). This 
nonphysical straight extension of the PL 
component to the low energy parts of the spectra is the main cause
for the required high values of $N_H$ in the MCD+PL fits, compared to 
those in the CMCD fits  (Tables~\ref{tab:fit-parameters} and
\ref{tab:evolution}).

\section{Comparisons with Independent Measurements}

In addition to the above self-consistency check of the X-ray spectral models,
we compare the inferred parameter values of the BHXBs with the independent
measurements to further the test. The key results of this 
comparison are included in Table~\ref{tab:keymeasure} and are discussed
in the following:

%\begin{itemize}
\noindent $\bullet$ {\bf X-ray Absorption}
%\item{
$N_H$ may be obtained more directly and accurately via 
the spectroscopy of neutral 
absorption edges of Oxygen and Neon (e.g., Schulz \etal~2002; Page \etal~2003).
This method is nearly independent of the overall continuous spectral shape 
and any other source properties. Using the dispersed spectrum of an 
{\sl XMM-Newton} observation, \citet{pag03} inferred
$N_H = 0.38_{-0.07}^{+0.08}\times 10^{21}~\rm{cm}^{-2}$ for LMC~X--3,
assuming the interstellar medium abundance of \citet{wilms00}.
We have also adopted this assumption by using the 
absorption model {\sl TBabs} in XSPEC. 
The $N_H$ values from the CMCD models and from the X-ray absorption edge
measurement agree with each other within the quoted error 
bars, whereas the value from the MCD+PL model is 
significantly higher (Table~\ref{tab:keymeasure}). 

\noindent $\bullet$ {\bf Disk Inclination Angle}
%\item{
The upper limits on $\theta$ from the spherical CMCD model are 
consistent with 
the values obtained from the optical observations 
(Table~\ref{tab:keymeasure}). There is a simple reason why 
$\theta$ can be constrained in the spherical CMCD model: the 
Comptonized flux is nearly isotropic and barely affected by the disk 
inclination, whereas the observed strength of the soft disk 
component is proportional to cos($\theta$). This mostly geometric
effect is strong when $\theta$ is large; e.g., no radiation come directly 
from an edge-on disk. Therefore, the upper limit can be constrained 
reasonably well, which is especially important for estimating the BH mass 
(cf. Eqs.~\ref{equ:cmcd} and \ref{equ:mcd}). In the slab-like CMCD model,
since both of the observed soft and hard components strongly depend upon 
the system inclination angle (e.g., Sunyaev \& Titarchuk 1985; see also 
the Fig.~2 in Paper I), $\theta$ in principle can be well constrained in 
this model. For LMC~X--3, the PL-like component only contributes 
a small portion of the total flux, and the constraint of $\theta$ mainly
from the soft component, as in a spherical CMCD model. 
For LMC~X--1, the PL-like component contributes
significantly to the total flux. 
The slab-like corona configuration becomes problematic, giving 
the inconsistent $\theta$ constraints.

\noindent $\bullet$ {\bf Black Hole Mass}
The mass of each putative BH may be estimated as 
$M = c^2fR_{in}$/$G\alpha$, where 
$\alpha = 6$ or 1 for a non-spin or extreme spin BH),
$f R_{in}$ is the inner disk radius, assumed to be the
same as the radius of the last marginally stable orbit around the BH, and the 
factor $f$ [depending on the system inclination, 
0.94 for LMC X--1 and 1.12 for LMC X--3; we adopt a new value of 
$\eta=0.41$ (see \S1) derived from \citet{kub98}, compared to the old values
used $\eta=0.7$ used in \citet{zhang97} and in Paper II] 
includes various corrections 
related to the spectral hardening, special 
and general relativity effects (e.g., Cunningham 1975; 
Zhang, Cui, \& Chen 1997; Gierli\'{n}ski \etal 1999, 2001). 
Assuming no spin for the BHs, from the spherical CMCD results, 
we estimate the BH masses 
as 4.0(3.8--4.5), 4.19(4.17--4.21)~$M_{\odot}$ for 
LMC~X--1 and LMC~X--3, respectively. For LMC~X--1,
the value is  consistent with the result from the optical study,
whereas for LMC~X--3, the derived $M_{BH}$ is slightly smaller
(Table~\ref{tab:keymeasure}) which may suggest that the LMC~X--3
is a mild spin system.

%\end{itemize}

\section{Summary}

In this work, we have applied the MCD+PL model as well as our recently 
constructed CMCD model to 
BHXB systems LMC~X--1 and LMC~X--3, confronting the fitted parameters
with directly measured values. 
We have also tested two different corona geometric configurations.
The spherical configuration passes almost all the tests:
the effective hydrogen column densities, disk
inclinations, and the BH masses of the LMC~X--1 and LMC~X--3.
This consistency suggests that the CMCD model with a spherical corona provides
a reasonably good spectral characterization of BHXBs. The model offers
useful insights into physical properties of the Comptonization coronae 
and their relationship to the accretion process. 
In contrast, the slab-like CMCD model is problematic in describing 
the spectrum of LMC~X--1, in which the PL-like component contributes 
significantly. Similarly, the MCD+PL model, though generally providing a 
good fit to the spectra of BHXBs, could give misleading parameter values.
Although the tests conducted in this work are still very limited, they
have demonstrated the potential in discriminating among various models.

\section*{Acknowledgments }
We thank the anonymous referee for his/her insightful comments on our 
manuscript which helps us to improve the paper greatly. Y. Yao is also
grateful to Xiaoling Zhang and Yuxin Feng for their useful discussions.

\end{document}